\documentclass[3p,twocolumn]{elsarticle}

\newenvironment{hangref}
  {\begin{list}{}{\setlength{\itemsep}{4pt}
  \setlength{\parsep}{0pt}\setlength{\leftmargin}{+\parindent}
  \setlength{\itemindent}{-\parindent}}}{\end{list}}

\usepackage{graphicx}
\usepackage{graphicx,epsf} 
\usepackage{dcolumn}
\usepackage{bm}
\usepackage{latexsym}

\begin{document}

\title{An Efficient Implementation of the Robust Tabu Search Heuristic for Sparse Quadratic Assignment Problems}

\author{G. Paul}
\ead{gerryp@bu.edu}

\address{Center for Polymer Studies and Dept. of Physics, Boston
             University, Boston, MA 02215, USA}

\begin{abstract}

  We propose and develop an efficient implementation of the robust
  tabu search heuristic for sparse quadratic assignment problems.
  The traditional implementation of the heuristic applicable to all
  quadratic assignment problems is of $O(N^2)$ complexity per
  iteration for problems of size $N$.  Using multiple priority queues
  to determine the next best move instead of scanning all possible
  moves, and using adjacency lists to minimize the operations needed to
  determine the cost of moves, we reduce the asymptotic ($N \rightarrow
  \infty$) complexity per iteration to $O(N log N )$.  For practical
  sized problems, the complexity is $O(N)$.
\end{abstract}


\maketitle

\section{Introduction}

The quadratic assignment problem (QAP) is a combinatorial optimization
problem first introduced by Koopmans and Beckman (1957).  It
is NP-hard and is considered to be one of the most difficult problems
to be be solved optimally.  The problem was defined in the following
context: A set of $N$ facilities are to be located at $N$ locations.  The
distance between locations $i$ and $j$ is $D_{i,j}$ and the quantity of
materials which flow between facilities $i$ and $j$ is $F_{i,j}$.  The
problem is to assign to each location a single facility so as to minimize
the cost  
\begin{equation}C=\sum_{i=1}^N \sum_{j=1}^N  F_{i,j} D_{p(i),p(j)}.
\end{equation}
where $p(i)$ represents the  location to which  facility $i$ is assigned.  It will be  helpful to think of the $N$ facilities and the matrix of flows between them in graph theoretic
terms as a graph of $N$ nodes and weighted edges, respectively.

There is an extensive literature which addresses the QAP and is
reviewed in Pardalos et al. (1994), Cela (1998), Anstreicher (2003),
Loiola et al. (2007, and James et al. (2009a).  With the exception of
specially constructed cases, optimal algorithms have solved only
relatively small instances with $N \le 36$.  Various heuristic approaches
have been developed and applied to problems typically of size
$N\approx 100$ or less. One of the most successful heuristics to date
for large instances is {\it robust tabu search}, RTS, (Taillard
(1991).  The use of tabu search for the quadratic assignment problem
has been studied extensively (Drezner (2005), Hasegawa et al. (2000),
James et al, (2009a, 2009b), McLoughlin and Cedeno (2005), Misevicius
(2007), Misevicius and Ostreika (2007), Skorinkapov (1994), and Wang
(2007)).  Some of the best available algorithms for the solution of the QAP are the hybrid genetic algorithms that use tabu search as an improvement mechanism. (See  Drezner (2008)).  

Here we will consider the robust tabu heuristic applied to {\it
  sparse} QAP instances. That is, the number of non-zero entries in
the either the flow matrix and/or the distance matrix is of $O(N)$ as
opposed to $O(N^2)$. Without loss of generality we will assume the
flow matrix is sparse.  Many real world problems are sparse.  In fact,
this work was motivated by the study of random regular sparse graphs.  
These graphs are very robust to partitioning and collapse due to  
removal of nodes or edges.  We are interested in the problem of determining 
how to assign the nodes of such a graph to locations in a metric space such 
that the total edge length of the graph is minimized;  this problem maps 
directly to a quadratic assignment problem.

There has been some previous work on sparse quadratic assignment problems.  
Milos and Magirou (1995) developed a
Lagrangian-relaxation lower-bound algorithm for sparse problems and
Panos et al.  (1997) developed a version of their GRASP heuristic for
sparse problems. However, to the best of our knowledge, an efficient
implementation of the robust tabu heuristic for sparse QAP instances
has not been proposed.

\section{Background - the tabu heuristic}

The tabu heuristic for the quadratic assignment problem consists of repeatedly
swapping locations of two nodes.  A single iteration of the heuristic
consists of

\begin{itemize}

\item[(a)] Determining the move which most decreases the total cost.
  Under certain conditions (see Section 4), if a move which lowers the 
  cost is not available, a move which raises the cost is made. So 
  that cycles of the same moves are
  avoided, the same move is forbidden ({\it taboo}) until a
  specified later iteration; we call this later iteration the {\it
    eligible iteration} for a given move.  This eligible iteration is
  traditionally stored in a {\it tabu list} or {\it tabu table}.
\item[(b)] Making the move.
\item[(c)] Recalculating the new cost of all moves.

\end{itemize}

The process is repeated for a specified number of iterations.
Traditional implementations of robust tabu search require $O(N^2)$
operations per iteration. The complexity of $O(N^2)$ is achieved by maintaining  a matrix  containing the cost  $\Delta(p,u,v)$ of swapping $u$ and $v$ for all $u$ and $v$, given a current assignment $p$.

 The complexity of the each step above is
as follows:

\begin{itemize}
\item[(a)] $O(N^2)$ - all possible $N(N-1)/2$ moves are considered.  The cost of each move is retrieved from $\Delta(p,r,s)$
\item[(b)] $O(1)$ - the locations of the two swapped nodes are simply transposed.
\item[(c)] $O(N^2)$ - based on the following observations of Taillard (1991):  

Following Taillard (1991), starting from an assignment of facilities $p$ let the resulting assignment after swapping facilities $r$ and $s$ be $p'$.  That is:
\begin{eqnarray}
p'(k) &=& p(k) ~~~~~   k\ne r,s   \nonumber \\
p'(r) &=& p(s)                                \nonumber \\
p'(s) &=& p(r).
\end{eqnarray}
For a symmetrical matrix with a null-diagonal, the cost  $\Delta(p,r,s)$ of swapping $r$ and $s$ is:
\begin{eqnarray}
\Delta(p,r,s) &=&  \sum_{i=1}^N \sum_{j=1}^N (F_{i,j} D_{p(i),p(j)} - F_{i,j} D_{p'(i),p'(j)}) \nonumber \\
&=& 2 \sum_{k \ne r,s} (F_{s,k} -F_{r,k}) (D_{p(s),p(k)-D_{p(r)},p(k)}.
\label{eqfull}
\end{eqnarray}

To calculate $\Delta(p',u,v)$ for any $u$ and $v$ with complexity O(N) , we can use equation xx. For asymmetric matices or matrices with non-null diagonals, a slightly more complicated version of Equation (\ref{eqfull}) also of complexity O(N) is given by Burkhard and Rendl (1984) .

 To calculate $\Delta(p',u,v)$ in the case that the swapped facilities $u$ and $v$ are different from  $r$ or $s$,  we use the value $\Delta(p,u,v)$ calculated in the previous iteration  and find:
\begin{eqnarray}
\Delta(p',u,v) = \Delta(p,u,v)                                 
+2(F_{r,u} - F_{r,v} + F_{s,v} -F_{s,u})                \nonumber \\
\cdot (D_{p'(s),p'(u)} - D_{p'(s),p'(v)} + D_{p'(r),p'(v)} - D_{p'(r),p'(u)})
\end{eqnarray}

\begin{itemize}
\item[(i)] the cost of moves which do not involve the two nodes in the
  previous move can be calculated in time $O(1)$.  There are $O(N^2)$ of
  these moves.
\item[(ii)] The cost of moves which do involve the two nodes in the
  previous move must be calculated from scratch.  There are $O(N)$ of
  these moves and the complexity of calculating each is $O(N)$.
\end{itemize}

\end{itemize}
  
\section{Approach}

To reduce the complexity of step (a), instead of scanning all possible moves, 
we use multiple {\it priority queues} (PQs) to determine the best move. A priority queue is a data structure for maintaining a set of elements each of which has an associated value (priority).  A PQ supports the following operations:
\begin{itemize}
\item Insert an item
\item Remove an item
\item Return the item with the highest value
\end{itemize}
Priority queues are used to efficiently find an item with the highest value without searching through all of the items.  

The maximum complexity of PQ operations is $O(log N)$. We will see below that
there will be $O(N)$ insertions and deletions in the PQs for each
iteration so the asymptotic complexity of this step is reduced to
$O(N log N)$.  Furthermore, we will show that for problems of any
practical size, PQ operations are not the determinant of total
complexity.

The complexity to recalculate the cost of moves in step (c), can be
reduced to $O(N)$ as follows:
\begin{itemize}

\item As in the traditional robust tabu implementation, the cost of
  moves which do not involve the two nodes in the previous move can be
  calculated in time $O(1)$.  On average, there are $2 \langle k
  \rangle $ nodes which are connected to the two nodes in the previous
  moves, where $\langle k \rangle $ is the average degree (average
  number of nodes adjacent to a given node) of the graph corresponding
  to the flow matrix.  For each of these $2 \langle k \rangle $ nodes
  we must calculate the cost of $N-1$ possible moves.  Thus, the cost
  is $O(\langle k \rangle N)$.

\item The cost of moves which do involve the two nodes in the previous
  move must be calculated from scratch.  There are $O(N)$ of these
  moves and the complexity of calculating each is $O(\langle k
  \rangle)$ since the cost of a node, $u$, being in a specific
  location depends only on the on-average $k$ nodes adjacent to $u$.

\end{itemize}
Thus the complexity of step (c) is reduced to $O(N)$.

\begin{figure}[tbh]
\centerline{
\epsfxsize=7.0cm
\epsfclipon
\epsfbox{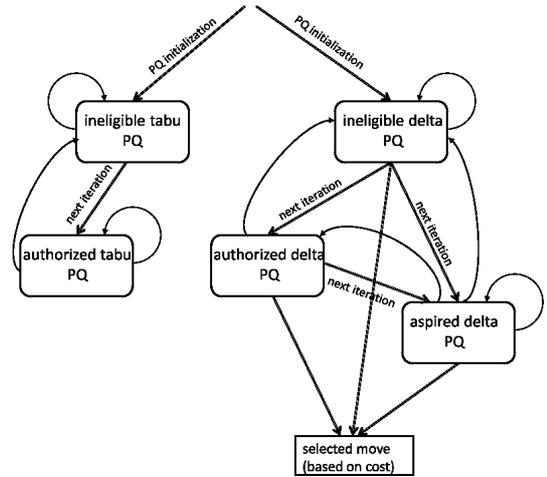}
}
\caption{Priority queues used in the implementation.}
\label{pqs}
\end{figure}

\section{Implementation}

To describe our implementation, we must first describe the rules 
for determining the next move of Taillard's
robust tabu heuristic (Taillard (1991)). The following definitions 
for the possible {\it state} of a potential move are useful:

\begin{itemize}
\item[(i)] If the current iteration is less than or equal to the
  eligible iteration, the move is {\it ineligible}.
\item[(ii)] If the current iteration is greater than the eligible
  iteration, the move is {\it authorized}.
\item[(iii)] If the current iteration minus an {\it aspiration
    constant} is greater than the eligible iteration the move is {\it aspired}.
\end{itemize}

The rules for determining the next move can then be stated as (Taillard (1991)):

\begin{itemize}
\item[(1)] If a move which decreases the lowest total cost found so
  far is available, the move which most decreases this total cost is chosen, independent of 
whether the move is ineligible, authorized or aspired.
\item[(2)] If no move meets criterion (1), the aspired move, if one is available,
  which most decreases the current total cost is chosen.
\item[(3)] If no moves meet criteria (1) or (2), the lowest cost
  authorized move is chosen.
\end{itemize} 

To implement these rules for sparse problems, we use two types of PQs: {\it delta} PQs which contain the cost delta for a
given move and {\it tabu} PQs which contain entries ordered by the
eligible iteration for the move.  The tabu PQs control the change of
state of a move.  The delta PQs determine the lowest cost move in each
state.  Five PQs are used:

\begin{itemize}

\item {\it ineligible tabu PQ} - This PQ contains moves, ordered by
  eligible iteration, which are in the ineligible state. This PQ
  allows us to efficiently determine when the state of a move can be
  changed to authorized.

\item authorized tabu PQ - This PQ contains moves, ordered by
  eligible iteration, which are in the authorized state.  This PQ
  allows us to efficiently determine when the state of a move can be
  changed to aspired.

\item ineligible delta PQ - This PQ contains moves, ordered by the
  cost of the move, which are in the ineligible state.  This PQ together 
  with the  two other delta PQs allows for
  efficient determination of the overall lowest cost move as required
  by rule 1.

\item aspired delta PQ -  This PQ contains moves, ordered by the
  cost of the move, which are in the aspired state. This PQ allows for
  efficient determination of the lowest cost aspired move as required
  by rule 2.

\item authorized delta PQ -  This PQ contains moves, ordered by the
  cost of the move, which are in the authorized state. This PQ allows
  us to determine the lowest cost authorized move as needed by rule
  3.

\end{itemize}

As illustrated in Fig. \ref{pqs}, moves are inserted and removed in
the PQs under the following circumstances:

\begin{itemize}

\item At initialization all moves are inserted into the ineligible PQs.

\item At the beginning of each iteration, any moves on the ineligible
  PQs which become authorized, because the iteration has increased
  by one, are moved from the ineligible PQs to the authorized
  PQs.

\item At the beginning of each iteration, any moves on the authorized
  PQs which become aspired, because the iteration has increased by
  one, are moved to the aspired PQ.
  
\item After each move, any move for which the move cost or eligible
  iteration has changed is removed from the PQ in which it is present
  and inserted in the appropriate PQ based on move state and move cost.
  (However, see lazy update discussion below.)

\end{itemize}
Using these PQs we obtain {\it exactly} the same results as the
traditional robust tabu implementation.

\section{Lazy Update}
We minimize the time updating PQs by performing {\it lazy updates}.
After a change in the eligible iteration or move cost, if the state is
changed or the value is increased, we update the PQs involved;
otherwise, we perform a lazy update and store the value in a data
structure associated with the move and only do the update in the PQ
when and if the move becomes the move with the smallest value in the
PQ.  This use of lazy updates significantly decreases the time spent
on PQ operations.

\begin{figure}[tbh]
\centerline{
\epsfxsize=7.0cm
\epsfclipon
\epsfbox{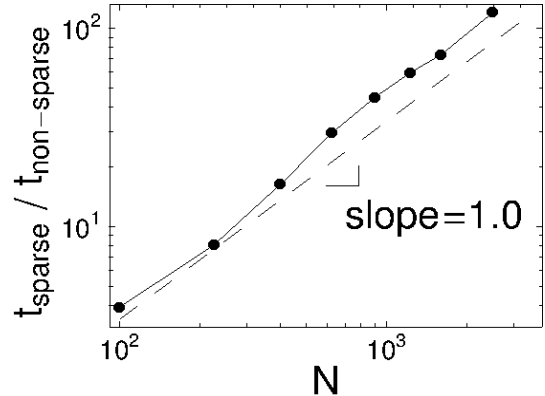}
}
\caption{The ratio of the time per iteration for the non-sparse implementation to the time per iteration for the sparse implementation versus the problem size, $N$. The slope of the plot is approximately 1.0 which is consistent with a reduction of $O(N)$ of the computational complexity}.
\label{ratio}
\end{figure}

\begin{figure}[tbh]
\centerline{
\epsfxsize=7.0cm
\epsfclipon
\epsfbox{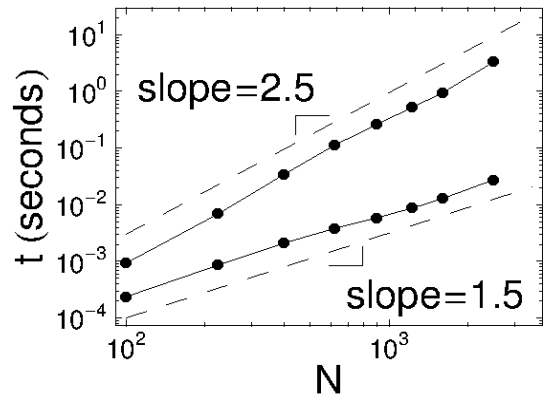}
}
\caption{The time per iteration for the non-sparse (upper plot) and sparse (lower plot) implementations.}
\label{comb}
\end{figure}

\begin{figure}[tbh]
\centerline{
\epsfxsize=7.0cm
\epsfclipon
\epsfbox{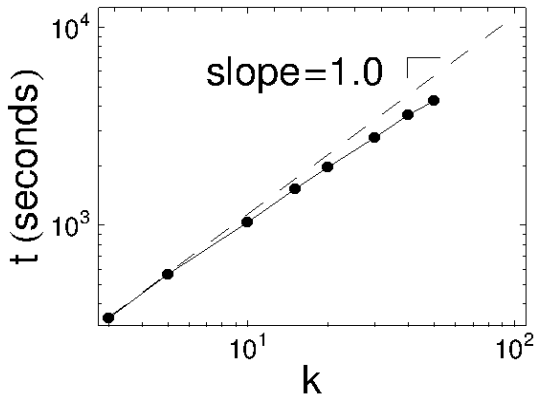}
}
\caption{The time per iteration using the sparse implementation for $N=400$ versus degree $k$.}
\label{tk}
\end{figure}

\section{Numerical Results}

We test our algorithm on instances with $N$ locations on a square grid
with a Euclidean metric.  The flow matrix for the $N$ facilities
corresponds to an adjacency matrix for a $k$-regular random graph;
that is, each facility has flows of value one to $k$ other random
facilities.  We run both the traditional non-sparse implementation and
our new sparse implementation.  The non-sparse implementation is the
code of Taillard (1991).  To implement the priority queues in the
sparse implementation, we use the complete balanced tree code of Marin
and Cordero (1995) .  We perform a minimum of $10^4$ iterations for
each instance.  

For $N=2500$, priority queue update operations consume just $2 \%$ of
the total time.  Assuming a $log N $ behavior for PQ
operations, it is not until $N$ becomes astronomically large that the PQ 
operations would take the majority of the time and 
the behavior of our implementation crosses over from
$O(N)$ to $O(N log N )$ complexity.  We thus expect $O(N)$ complexity for
our sparse implementation for any practical sized problems.

In Fig.  \ref{ratio}, for $k=3$, we plot the ratio $r$ of the time
per iteration of the non-sparse implementation to the time per
iteration of the sparse implementation versus $N$.  As
expected, the plot is consistent with $r \sim N^x $ where $x~ \approx
1.0$.  This reflects the reduction in complexity from $O(N^2)$ to
$O(N)$.  In Fig.  \ref{comb} we plot separately the time per iteration
for the original and the sparse implementations.  Consistent with Fig.
\ref{ratio}, the slopes on the log-log plot differ by 1 but the slopes
are 2.5 and 1.5, respectively, as opposed to the theoretical values of
2.0 and 1.0.  As explained in Paul (2007) and Saavedra and Smith
(1995), this is due to the finite size of processor cache
memory; as the problem size (and memory needed) increases, there are a
smaller percentage of cache hits causing slower operation.

In Fig. \ref{tk} we plot the time per iteration versus values of degree $k$.
The plot is linear with a slight deviation with increasing $k$.  This
deviation from linear is due to the fact that as $k$ increases
there is an increasing chance that a node $u$ will be connected to both
nodes involved in the previous move; however, the updated costs of moving
node $u$ must be calculated only once.

We also performed numerical experiments on a class of  problems known in the literature (see Drezner et al. (2005)).  These problems denoted as DRExx are sparse and are designed to be very difficult to solve with heuristics.  We obtained results similar to those for described above; performance is O(N).

\section{Summary}

For sparse quadratic assignment problems, we reduce the asymptotic
complexity per iteration of robust tabu search to $O(N log N)$ from
$O(N^2)$; for practical size problems, the complexity is reduced to
$O(N)$. Central to achieving this reduction is the use of multiple
priority queues and lazy updates to these queues.  The code which
implements our approach and test QAP instances used for this
paper are available as supplementary material.

\section*{Acknowledgment}
We thank Jia Shao for helpful discussions and the Defense Threat Reduction Agency (DTRA) for support.

\section*{References}

\begin{hangref}

\item Anstreicher, K., 2003. Recent advances in the solution of quadratic assignment problems. Mathematical Programming 97, 27-42.

\item Burkhard R. E. F. Rendl, 1984. A thermodynamically motivated simulation procedure for combinatorial optimization problems.  EJOR 17, 169-174,

\item Cela E., 1998. The Quadratic Assignment Problem: Theory and Algorithms. Kluwer, Boston, MA.

\item Cormen, T.H., Leiserson, C. E., Rivest R.L., Stein C., 2009.  Introduction to Algorithms. MIT Press. 

\item Drezner, Z. , 2002. Heuristic algorithms for the solution of the quadratic assignment problem.  Journal of Applied Mathematics and Decision Sciences 6, 163-173.

\item Drezner, Z., 2003. A new genetic algorithm for the quadratic assignment problem.  INFORMS Journa of Computing 15, 320-330.

\item Drezner, Z., 2005. Compounded genetic algorithms. Operations Research Letters 33, 475-480.

\item Drezner, T., Drezner, Z., 2006. Gender-specific genetic algorithms. INFOR 2006 44 117-126.

\item Drezner, Z., 2005. A distance based rule for removing population members in genetic algorithms. 4OR 3, 109-116.

\item Drezner, Z. Marcoulides G.A.,2006.  Mapping the convergence of genetic algorithms. Advances in Decision Science 
 2006 Article ID 70240, 16 pages.

\item Drezner, Z., 2005. The extended concentric tabu for the quadratic assignment problem. European Journal of Operational Research 160, 416-422.

\item Drezner, Z., Hahn,P.M., Taillard \`E. D., 2005. Recent Advances for the quadratic assignment problem with special emphasis on instances that are difficult for meta-heuristic methods. Annals of Operations Research 139, 65-94. 

\item Drezner, Z. 2008.  Extensive experiments with hybrid genetic algorithms for the solution of the quadratic assignment problem. Computers and Operations Research, 35, 717-736.

\item Drezner Z. and Marcoulides G., 2009.   On the Range of Tabu  Tenure in Solving Quadratic Assignment Problems   Recent Advances  in Computing and Management Information  Systems.  In: Recent Advances in Computing and Management Information  Systems, 157-169 , P. Petratos and G. A. Marcoulides (Eds),  ATINER.

\item Hasegawa, M., Ikeguchi,T., Aihara, K., 2000. Exponential and chaotic neurodynamical tabu searches for quadratic assignment problems.
CONTROL AND CYBERNETICS 29, 773-788.   

\item James, T., Rego, C., Glover, F., 2009a. Multistart Tabu Search  and Diversification Strategies for the Quadratic Assignment
  Problem. IEEE Tran. on Systems, Man, and Cybernetics  PART A: SYSTEMS AND HUMANS 39, 579-596.

\item James, T., Rego, C., Glover, F., 2009b. A cooperative parallel tabu search algorithm for the quadratic assignment problem. 
European Journal of Operational Research 195, 810-826.

\item Koopmans T., Beckmann, M., 1957. Assignment problems and the location of economic activities. Econometrica 25, 53-76.

\item Loiola,E.M., de Abreu, N.M.M., Boaventuro-Netto, P.O., Hahn, P., Querido, T., 2007. A survey for the quadratic assignment problem. European Journal of Operational Research 176, 657-690.

\item Mar\`in, M., Cordero P., 2007. An empirical assessment of priority queues in event-driven molecular dynamics simulation, Computer Physics Communications 92, 214-224.
 
\item McLoughlin, J. F., Cedeno, W., 2005. The Enhanced Evolutionary Tabu Search and its application to the Quadratic Assignment Problem.
GECCO 2005: Genetic and Evolutionary Computation Conference 1, 975-982. 

\item Milis, I. Z., Magirou V. F., 1995. A Lagrangian-relaxation algorithm for sparse quadratic assignment problems. Operations Research Letters. 17, 69-76.  

\item Misevicius, A., 2007. A tabu search algorithm for the quadratic assignment problem. Computational Optimization and Applications 30, 95-111.  
 
\item Misevicius,A., Ostreika, A., 2007. Defining tabu tenure for the
  quadratic assignment problem. Information Technology and Control.
  36, 341-347.

\item Pardalos, P.M., Rendl, F., Wolkowicz, H., 1994. The quadratic
  assignment problem: A survey and recent developments.  In: Pardalos,
  P.M., Wolkowicz, H., (Eds.), Quadratic Assignment and Related
  Problems. DIMACS Series on Discrete Mathematics and Theoretical
  Computer Science 16, Amer. Math. Soc., Baltimore, MD.  1-42.

\item Pardalos, P. M., Pitsoulis, L. S., Resende, M. G. C., 1997. Algorithm 769: Fortran subroutines for approximate solution of sparse quadratic assignment problems using GRASP. ACM Transactions on Mathematical Software (TOMS) 23, 196-208.

\item Paul, G., 2007. A Complexity $O(1)$ priority queue for event driven
molecular dynamics simulations. Journal of Computational Physics 221, 615-625. 

\item R.H. Saavedra, Smith, A.J., 1995.  Measuring cache and TLB
  performance and their effects on benchmark run times. IEEE
  Transactions on Computers 44, 1223-1235.
 
\item Skorinkapov, J., 1994. Extensions of a tabu search adaptation to
  the quadratic assignment problem. Computers and Operations Research
  121, 855-865.

\item Taillard, E., 1991.  Robust taboo search for the quadratic
  assignment problem. Parallel Computing 17, 443-455; code available
  at:http://mistic.heig-vd.ch/taillard/codes.dir/tabou\_qap.cpp.
 
\item Wang, J. C., 2007. Solving Quadratic Assignment Problems by a
  Tabu Based Simulated Annealing Algorithm.  ICIAS 2007: International
  Conference on Intelligent and Advanced Systems Proceedings, 75-80

\end{hangref}

\end{document}